\documentclass[prb,superscriptaddress,nofootinbib]{revtex4}

\usepackage{amsmath}
\usepackage{amsfonts}       
\usepackage{graphicx}
\usepackage{dcolumn}
\usepackage{epsfig}
\usepackage{bm}
\usepackage{array}
\usepackage{hyperref}
\hypersetup{
colorlinks=true,
citecolor=blue,
linkcolor=blue,
urlcolor=blue,
}
\usepackage{xcolor}			

\usepackage{graphicx}
\usepackage{amsmath}
\usepackage{amssymb}
\usepackage{bm}           
\usepackage{bbm}
\usepackage[english]{babel}
\usepackage{setspace}
\usepackage{wasysym}
\usepackage{adjustbox}
\usepackage{framed}
\usepackage{mdframed}
\usepackage{latexsym} 
\usepackage{pdflscape}   
\usepackage{multirow}    
\usepackage{booktabs}    
\usepackage{calc}        
\usepackage{bbold}

\usepackage{calrsfs}								
\DeclareMathAlphabet{\pazocal}{OMS}{zplm}{m}{n}	

\definecolor{DeepSkyBlue}{RGB}{0,104,139}
\colorlet{MySky}{white!40!blue}
\colorlet{MyViolet}{red!45!blue}
\colorlet{MyBlue}{black!40!blue}
\colorlet{MyRed}{black!40!red}
\colorlet{MyOrange}{red!70!yellow}
\colorlet{MyGreen}{black!60!green}
\colorlet{MyBrown}{black!70!brown}
\colorlet{MyGray}{black!60!white}

\newcommand{\be}{\begin{equation}}
\newcommand{\ee}{\end{equation}}

\newcommand{\beq}{\begin{eqnarray}}
\newcommand{\eeq}{\end{eqnarray}}

\begin{document}

\title{Integer Factorization by Quantum Measurements}

\author{G. Mussardo}
\affiliation{SISSA, Via Bonomea 265, I-34136 Trieste, Italy.}
\affiliation{INFN, Sezione di Trieste, Via Valerio 2, I-34127 Trieste, Italy}

\author{A. Trombettoni}
\affiliation{Dipartimento di Fisica, Universit\'a di Trieste, Strada Costiera 11, I-34151 Trieste, Italy}
\affiliation{INFN, Sezione di Trieste, Via Valerio 2, I-34127 Trieste, Italy}

 

\maketitle

{\bf Quantum algorithms are at the heart of the ongoing efforts to use quantum mechanics to solve computational problems unsolvable on ordinary classical computers \cite{Nielsen2010,Preskill,Pittenger}. Their common feature 
is the use of genuine quantum 
properties such as entanglement and superposition of states \cite{HorodeckiRMP2009}. 
Among the known quantum 
algorithms, a special role is played by the Shor algorithm \cite{Shor1994,Shor1995}, i.e. a polynomial-time quantum algorithm for integer factorization, with far reaching potential applications in several fields, such as cryptography \cite{Ekert}. For an integer $N$ of the order of $2^n$, i.e. with $n$ digits,  
the Shor algorithm permits its factorization in (order of) $n$ steps. This results in an exponential gain in computational efficiency with respect to the best known classical algorithms.  Here we present a different algorithm for integer factorization based on another genuine quantum property:
quantum measurement \cite{Peres1993,Shankar1994,Sakurai2020}.
In this new scheme, the factorization of the integer $N$ is achieved in a number of steps equal to the number of its prime factors, referred to as $k$
-- e.g., if $N$ is the product of two primes, 
two quantum measurements are enough, regardless of the number of digits $n$ of the number $N$. 
Since $k$ is the lower bound to the number of operations one can do to factorize a general integer, then one sees that a quantum mechanical setup can saturate such a bound. Once established this, we discuss how the algorithm can physically be ran. We argue that one needs a single-purpose device where quantum measurements of an observable with assigned spectrum can be performed. The preparation from scratch of this device requires the solution, once for all and not for each factorization operation, 
of $\sim 2^{n}$ differential equations, a task that with a quantum computer can be accomplished in $n$ steps. }

\vspace{3mm}
\noindent
\textit{Introduction. }
Recent progress in the implementation of quantum devices has led to the experimental demonstration of some instances of quantum advantage. This happens when a specific computational problem may be solved faster and more efficiently on quantum processors rather than a classical computer \cite{Wu2021}. To achieve this goal the quantum processor must have an architecture made at least of several tens of qubits and long enough decoherence times. 

A notable example, from a historical and conceptual point of view, of a clear quantum advantage is provided by the Shor algorithm \cite{Shor1994,Shor1995}. This algorithm indicates how to solve efficiently on a quantum computer the long-standing problem of finding the prime factors of an integer number $N$. Assuming that such a number $N$ is of order $2^n$, the Shor algorithm exploits in an ingenious way the implementation of the discrete Fourier transform on $n$ qubits. To date, its validity has been shown with the factorization of a small numbers (the present computational bottleneck being the quantum modular exponentiation).  The factorization of the number, $15 = 3 \times 5$, was 
done using $7$ qubits with an NMR implementation of a quantum computer \cite{Vandersypen2001}. Similar demonstrations were performed using photonic \cite{Lu2007,Lanyon2007} and solid-state qubits \cite{Lucero2007}, while in 2012, with the $n$ qubits control register replaced by a single qubit recycled $n$ times, it was achieved the factorization of the integer $21 = 3 \times 7$ \cite{Martin2012}. Despite their simplicity, these examples nevertheless provide a {\em proof of principle} realization of the algorithm. 

In this paper we present a different route for integer factorization, based on an algorithm which exploits another genuine quantum property: projective 
quantum measurement \cite{Peres1993,Shankar1994,Sakurai2020}. As it is well known from quantum mechanics axioms,  if a physical system is in a normalised state $|\psi \, \rangle$, a measurement of an observable $\hat{O}$ will yield one eigenvalue $\alpha$ of its spectrum with probability $\mid\langle \psi \mid \alpha \rangle\mid^2$, where $|\alpha \,\rangle$ is the normalised eigenfunction corresponding to the eigenvalue $\alpha$
\begin{equation}
\hat{O} \,|\alpha \,\rangle\,=\,\alpha \,|\alpha \, \rangle \,.
\end{equation}
As a result of the measurement, the system state will change from $|\psi \,\rangle$ to $|\alpha \,\rangle$. 
For problems related to number theory, 
interesting spectra to consider are: (a) the natural numbers, corresponding to the 
Hamiltonian of an harmonic oscillator \cite{Shankar1994,Sakurai2020}; (b) the primes \cite{GMscattering,holographic}; and (c) the logarithm of the primes \cite{mack10,Weiss,Schleich1,Schleich2,Schleich,MTZ}. Employing such spectra, one may translate number theory problems in quantum physical settings. As an example of this general philosophy, in this paper we show that with a suitable choice of the operator $\hat{O}$ is possible to determine the prime factors of an integer number $N$ by making a finite set of quantum measurements.

The layout of this article is as follows: after a brief reminder on classical algorithms on primality and factorization of integers, 
we present our quantum algorithm for the factorization problem. This algorithm is discussed in terms of the familiar setting of Schr\"{o}dinger Hamiltonian  
although, as shown in the Supplementary Materials, it also admits a digital implementation. The computational costs of Shor's and our algorithm 
are thoroughly discussed in the last part of the paper.   
In the Supplementary Materials we discuss the digital implementation of our algorithm and a {\em gedanken experiment} able to achieve, in principle, a projective measurement of quantum system eigenenergies.  

\vspace{5mm}
\noindent
\textit{Classical primality tests and factorization algorithms. } 
The fundamental theorem of arithmetic states that every natural number $N$ greater than $1$ is either a prime number $p_k$ or can be represented as a product of prime numbers
\begin{equation}
N = p_{a_1}^{\alpha_1} p_{a_2}^{\alpha_2} \ldots p_{a_k}^{\alpha_k} \,\,\,,
\label{primedecomposition1}
\end{equation}
where $p_{a_1} < p_{a_2} \cdots < p_{a_k}$ are $k$ ordered primes and $\alpha_i \geq 1$ their multiplicity. Hence, prime numbers may be regarded as the atoms of arithmetic but, in contrast with the finitely many chemical elements, the number of primes is instead infinite, as shown by a classic argument by Euclid dated more than 2000 years ago. The appearance of prime numbers along the integer sequence is completely unpredictable. However, their coarse graining properties, and in particular how many prime numbers there are below any real number $x$, are aspects which can be controlled with remarkable precision. In other words, while there is no known simple function $f(n)$ which gives the $n$-th prime number $p_n$ (and the actual determination of prime numbers can only be done by means of the familiar Eratosthenes's sieve \cite{Ribenboim,Schroeder,Zagier,Granville,Rose}), we have instead perfect knowledge of the inverse function $\pi(x)$ which counts the number of primes below the real number $x$ \cite{Hardy,Apostol,Tao,Ore,Ribenboim,Schroeder,Zagier,Granville,Rose,theorem1,theorem2,theorem3,theorem4,Riemannorig,Edwards,Borwein}. Such a function was 
exactly determined by Riemann (see, for instance \cite{Edwards}): it has a  staircase behaviour (since it jumps by $1$ each time $x$ crosses a prime), but becomes smoother and smoother for increasing values of $x$, and its asymptotic behaviour is constrained by the ``Prime Number Theorem'' \cite{theorem1,theorem2,theorem3,theorem4}stating that  
$\lim_{x\to\infty} \frac{\pi(x) \ln x}{x} \,=\,1
$.
Notice that $p_n \,=\, \pi^{-1}(n)$. Hence, inverting at the lowest order the function $\pi(x)$, one gets the following scaling law for the $n$-th prime number:  
$p_n \simeq n \,\log n$.

Let's first discuss the primality test. How can one tell whether an integer $N$ is prime? What if the number has hundreds or thousands of digits? This question may seem abstract or irrelevant, but primality tests are performed every time to make online transactions secure.
Given an integer $N$, determine whether or not $N$ is a prime number constitutes a primality test. 
The naive way to check the primality of an integer $N$ is to divide it by any prime number between $2$ and $\sqrt{N}$. Assuming we express the number $N$ in binary basis, if $N$ is of order $2^d$, the number of operations ${\mathcal D}$ of this naive primality algorithm scales exponentially with the number of digits, i.e. ${\mathcal D} \sim 2^{d/2}$. 
Many different algorithms have been proposed, both deterministic or probabilistic nature (see, for instance \cite{Gillen}), to reduce the complexity of the primality test. The final answer was given in 2002 \cite{Agrawal} in terms of a deterministic protocol (the AKS test) whose complexity scales as \cite{Agrawal}
${\mathcal D}_{AKS} \sim (\log N)^{7.5}$.

Let's now imagine that the primality test outputs that $N$ is not a prime. Then, looking at (\ref{primedecomposition1}), 
how to determine the prime factors $p_{a_1} \ldots p_{a_k}$ of the integer $N$? This is the question addressed by any factorisation algorithm. Let $N$ be a number of $n$-bit digits: currently there is no classical factorization algorithm whose complexity scales ${\mathcal D} \sim {\cal O}(n^{\alpha})$ for some constant $\alpha$. 
Although neither the existence nor non-existence of such algorithms has been proved, it is generally believed that they do not exist and hence that the problem is not in the class of Polynomial Time Algorithms. If so, then the problem is clearly in class NP, although it is not certain whether it is or is not in the class of NP-complete problems \cite{fact1,fact2,fact3}. 
The best classical algorithm known is the so-called general number field sieve (GNFS) \cite{fact2} whose complexity for a number $N$ scales as $\exp (\ln N)^{1/3}$.

\vspace{3mm}
\noindent
\textit{The factorization algorithm by quantum measurements. }
In the following we discuss how to devise a very efficient factorisation algorithm using
quantum mechanical measurements in a way that the number of steps is finite and does not scale with the $n$ digits of $N$.
In our presentation the operator $\hat{O}$ is the guise of the familiar Hamiltonian $\hat{H}$, but the reader is of course 
free to substitute the operator $\hat{H}$ with any other hermitian
operator with the appropriate spectra. Let assume that the Hamiltonian $\hat{H}$ is made up of two Hamiltonians $\hat{H}_1$ and $\hat{H}_2$ which commute to each other:
\be
\hat{H}\,=\, \hat{H}_1 + \hat{H}_2 \,, \,\,\,\,\,\,\,\,\,\,
    [\hat{H}_1, \hat{H}_2] = 0 \,\,\,.
\ee
We choose $\hat{H}_1$ to have $M$ eigenvalues given by the logarithms of the primes, where $M = 2^d$ is a fixed cut-off (which however can be increased as desired)
\begin{equation}
E_n^{(1)} \,=\,  \log p_n \,;
\hspace{2mm}
 n=1, 2, 3, \cdots 2^d \,.
 \label{logpp}
\end{equation}
The corresponding eigenfunctions will be denoted as $|\log p_n \rangle$.

On the other hand, the second term $\hat{H}_2$ in the Hamiltonian $\hat H$ is chosen to have the eigenvalues given by the logarithms of the integers, up to the same cut-off $M$, i.e.  
\begin{equation}
E_m^{(2)} =\log  m\,; 
\hspace{2mm}
 m = 2, 3, \ldots 2^d\,.
 \label{lognn}
\end{equation}
The corresponding eigenvectors will be denoted as $|\log m\rangle$. We will discuss below how such Hamiltonians $H_1$ and $H_2$ can be explicitly realised in a laboratory by means of spatial light laser modulators, as it was recently done for an Hamiltonian having the prime numbers as quantum spectrum \cite{holographic}.    

The generic eigenfunctions of $\hat{H}$ are then given by 
\begin{equation}
|\log p_n \rangle \, |\log m \rangle \,,
\end{equation}
which correspond to the eigenvalues  $E_{n,m} \,\equiv \, \log p_n + \log m$.

The factorization problem of a natural number $N$ consists of finding the primes entering its decomposition (\ref{primedecomposition1}).
In order to do so, our protocol consists of the 
following steps. 

\begin{enumerate}
\item[1.] Take initially the logarithm of the number $N$ to be factorized, promote it to be an eigenvalue of the Hamiltonian $\hat{H}$ and prepare the initial state $|\log N \rangle$.

\item[2.] For an integer $N$ -- as the one given in Eq.(\ref{primedecomposition1}) -- made of $k$ distinct primes, the corresponding energy level of $\hat{H}$ is $k$-fold degenerate, i.e. the degeneracy of the level depends {\em only} on the number of distinct primes present in $N$ and not on their multiplicities. Indeed, $\log N$ can be written in the following $k$ different ways 
\begin{eqnarray}
\log N & \,=\, & \log p_1 + \log \left(p_1^{a_1 -1} p_2^{a_2} ...p_{k-1}^{a_{k-1}} p_k^{a_k} \right)\,\equiv \, \log p_1 + \log \tilde N_1 \nonumber \\
\log N & \,=\, & \log p_2 + \log \left(p_1^{a_1} p_2^{a_2-1} ..p_{k-1}^{a_{k-1}} p_k^{a_k}\right) \,\equiv \, \log p_2 + \log \tilde N_2 \nonumber \\
\cdots & 
& \\
\log N & \,=\, & \log p_{k} + \log \left(p_1^{a_1 } p_2^{a_2} ...p_{k-1}^{a_{k-1}} p_k^{a_k  - 1} \right) \,\equiv \, \log p_k + \log \tilde N_{k} \nonumber 
\end{eqnarray}
where 
$\tilde N_a \,\equiv \, N/p_a$ 
is the integer obtained dividing the original number $N$ by one of its prime factor $p_a$. 

\item[3.] Hence, the generic state of the $k$-th degenerate manifold with energy $\log N$ (here simply denoted as $ | \log N \rangle $) admits the expansion 
\begin{equation}
| \log N \rangle \,=\, \sum_{a=1}^k c_a |\log p_a \rangle |\log N_a \rangle 
\hspace{5mm}
,
\hspace{5mm}
\sum_{a=1}^k |c_a|^2 \,=\,1 
\label{genericstate}
\end{equation}
For a generic state, we can assume that all coefficients of this expansion are different from zero. Their values are actually not essential for the running of the algorithm, 
so one may assume to be randomly distributed, as would occur in the case of a random initial state preparation.
\item[4.] After the state (\ref{genericstate}) is prepared, measure $\hat{H}_1$. With all coefficients $c_a$ different from zero, the output will be the logarithm of 
one of the primes present in $N$, say $\log p_b$, with probability $|c_b|^2$. 
\item[5.] Once the result of this measurement is known, divide the original number $N$ by the prime $p_b$ identified by the output of the $\hat{H}_1$ measurement. In this way 
one obtains the lower integer $\tilde N_b=N/p_b$. Then, start over, taking $\tilde N_b$ as the integer to be factorized. The procedure will halt after a number of iterations $l$ equal to the total number of primes present in $N$, i.e. 
 \begin{equation}
 I \,=\, \sum_{l=1}^k \alpha_l \,\,\,.
 \end{equation}
\end{enumerate}
Notice that the number of quantum measurements can be made equal to the number of distinct factors 
substituting the previous point $5.$ with this new one
\begin{enumerate}
\item[5.'] Once the result of this measurement is known, divide the original number $N$ by the prime $p_b$ identified by the output of the $\hat{H}_1$ measurement.  In this way 
one obtains the lower integer $N/p_b$. Use a classical computer to continue to divide for $p_b$ till the obtained number is not longer divisible for $p_b$. In this way one obtains the multiplicity $\alpha_b$ associated to the factor $p_b$. Then, start again the procedure, taking $\tilde N_b=N/p_b^{\alpha_b}$ as the integer to be factorized. 
\end{enumerate} 

\begin{figure}[t]
 \includegraphics[width=0.4\textwidth]{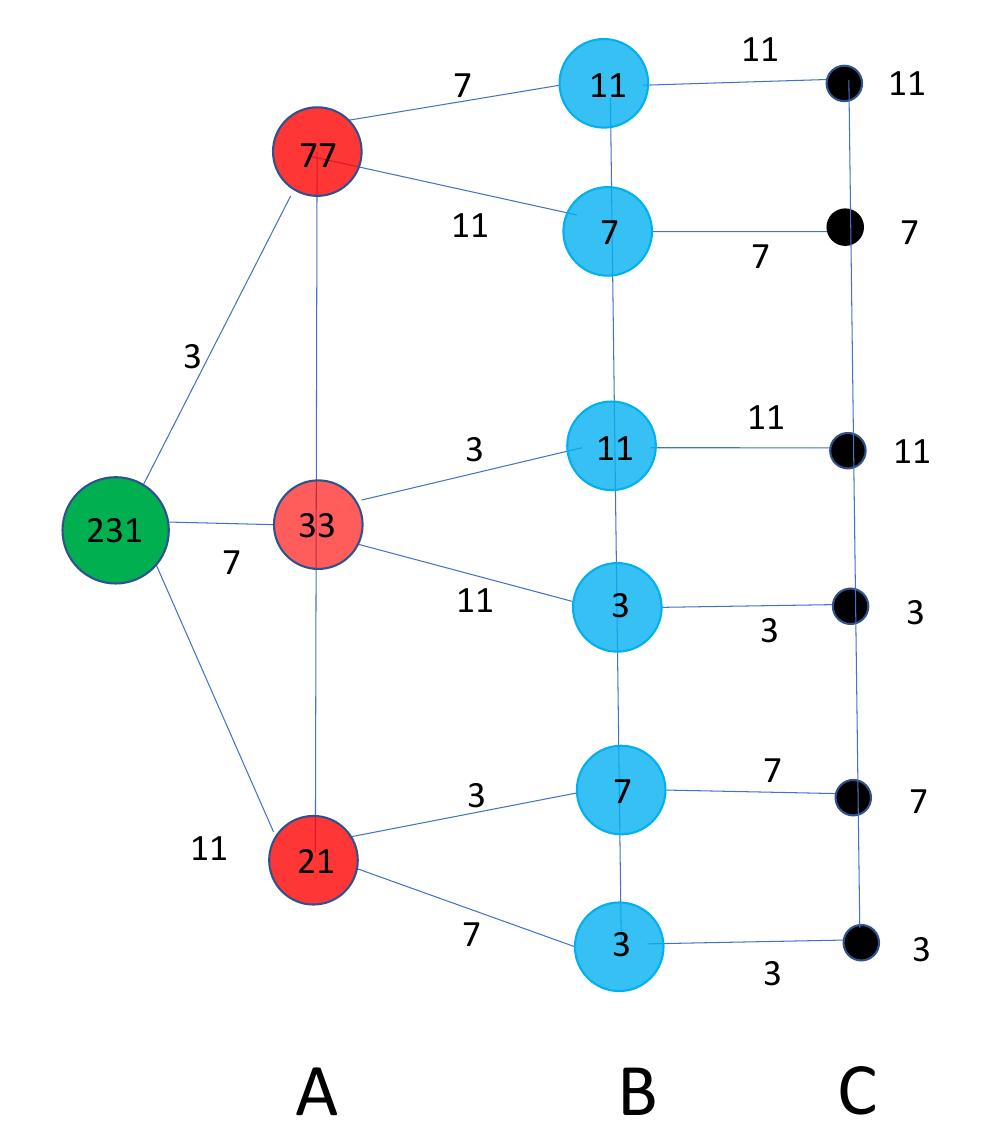}
    \caption{The graph associated to the various quantum measurements of $\hat{H}_1$ which pin down the different prime factors of $N$. In the example shown in the figure, the number to factorize is $N= 231 = 3 \times 7 \times 11$. For simplicity, along the links of the graph, we indicate the primes (rather than their logarithm) which are the outputs of the three stages $A, B, C$ measurements of $\hat{H}_1$. One gets a complete factorisation independently of the path associated to the chain of $\hat{H}_1$ measurements.}
    \label{graphmeasurement}
\end{figure}

\noindent
Four significant remarks are in order:
\begin{itemize}
\item A primality test can be immediately implemented by performing a single quantum measurement of $\hat{H_1}$ on the initial state $|\log N \rangle$. If the system collapses (remains) in itself, $N$ is a prime. 
\item A discussion on the preparation of the intial state is in the Supplementary Materials. 
\item If the multiplicity of the last factor, say $p_k$, to be factorized is $1$, then the last operation with the measurement of $\hat{H}_1$ actually amounts to apply the identity operator. If the multiplicity of different from $1$, i.e. $\alpha_k>1$, then $\alpha_k$ subsequent quantum measurements of $\hat{H}_1$ will produce each time with probability $1$ the same eigenstate $|\log p_k \rangle$.

\item The successful implementation of the algorithm is guaranteed independently of the $\hat{H}_1$ outputs obtained at the various stages of the algorithm. It is a {\em all roads lead to Rome} procedure. As shown in Figure  \ref{graphmeasurement}, there are possible branches resulting from successive measurements of $\hat{H}_1$. 
Imagine, for instance, that we want to factorize the number $N = 231$, whose prime decomposition is given by $N = 3 \times 7 \times 11$. As a result of the first 
measurement, we could have one of three possibie outputs: $\log 3$, $\log 7$ or $\log 11$. 
\begin{enumerate}
\item If the first output is $\log 3$, the next integer to be factorized is $\tilde N_3 = 77$ and the next measurement of $H_1$ starting from the (log) of this number, can give, as output, either $\log 7$ or $\log 11$. Once this last measurement is made, the next number is uniquely determined, thus arriving to the complete factorization of the number $N=231$. It is easy to see that the same conclusion will be reached if there is a different initial output. 
\item  If the first output is $\log 7$, the next integer to be factorized is $\tilde N_7 = 33$ and the next measurement of $H_1$ starting from the (log) of this number, can give, as output, either $\log 3$ or $\log 11$. Once  this last measurement is made, the next number is uniquely determined, thus arriving to the complete factorization of the number $N=231$. 
\item  If the first output is $\log 11$, the next integer to be factorized is $\tilde N_{11} = 21$ and the next measurement of $H_1$ starting from the (log) of this number, can give, as output, either $\log 3$ or $\log 7$. Once this last measurement is made, the next number is uniquely determined, thus arriving at the complete factorization of the number $N=231$. 
\end{enumerate}
\end{itemize} 

The key feature of this algorithm is the projective quantum measurements of hermitian operators, such as the Hamiltonians we have employed. This leads to an algorithm with the least possible number of operations, equal to $k$, relative to the factorization of an integer made of $k$ primes, see below for a discussion of this point. 

Given that quantum mechanics is based on linear algebra while the mathematical problem relative to factorization involves products, Hamiltonians with logarithm of the primes as spectrum are very natural to study and have been also used before \cite{Weiss,Schleich1,Schleich2, Schleich}. In more detail, a thermally isolated non-interacting Bose gas loaded in a one-dimensional potential with logarithmic energy eigenvalues was discussed in \cite{Weiss}, where asymptotic formulas for factorising products of different primes were provided. Time-dependent perturbation in (possibly multiple copies of) a potential having logarithms of the primes as eigenvalues was instead proposed in \cite{Schleich1,Schleich2,Schleich}, where the number $N$ to be factorized is encoded in the frequency of a sinusoidally modulated interaction acting on the ground-state of the potential. In this approach, if the integer is a product of $k$ primes, one needs to prepare an ensemble of $k$
identical systems each with an energy spectrum given by the logarithm of the primes \cite{Schleich}. Recently, a paper discussed the possible use of an Hamiltonian having as eigenvalues the logarithm of the primes for search algorithm \cite{arm}.

Our approach differs from the one discussed in \cite{Schleich} for two key features: first, the presence of $\hat{H}_2$ in our Hamiltonian and, secondly, the use of quantum measurements. Employing our Hamiltonian, made of a piece relative to the logarithm {\em and} another one relative to the logarithm of the integers, it is possible to factorize an integer $N$ without knowing a-priori the number $k$ of distinct primes (and their multicplicities) this number $N$ is made of. 

Concerning the quantum measurement of $\hat{H}_1$, one could use the von Neumann scheme \cite{vonN}, which consists of setting up a coupling with a test particle of the form $\hat{p} \, \hat{O}$, where $\hat{p}$ is the momentum operator of the test particle and $\hat{O}$ the operator to be measured. 
In our case, those quantum measurements can be performed in a slightly different way following the gedanken experiment discussed in the Supplementary Materials.

\vspace{3mm}
\noindent
\textit{Quantum Potentials. } 
Let's now discuss how to implement in a laboratory the algorithm described in the previous section. One possibility is to adopt a \textit{digital} route: this consists of implementing a discrete quantum register of two-level systems, such spin-$1/2$. This can be done with a quantum ladder, where in the first leg one constructs the  Hamiltonian $\hat{H}_1$ whose eigenvalues are the logarithms of the primes, while in the second leg the Hamiltonian $\hat{H}_2$ whose eigenvalues are the logarithms of the integers. Both legs should have $d$ spins. The eigenvalues should be respectively the logarithms of the first $2^d$ primes and the logarithms of the first $2^d$ integers. Given the current trend of quantum computing technology, this seems the simplest thing to do. However, as discussed in more detail in the Supplementary Materials, this digital implementation of our algorithm has the following bottleneck: while realising $\hat{H}_2$ is an easy task (since it consists of fixing a number of couplings which is linear in $d$), realizing $\hat{H}_1$ needs instead to tune an exponential number of long-range, many-body couplings between the qubits, and it is fair to say that this obstacle seems unlikely to be overcome soon, at least with the present technology. 
Even more, the tuning of this exponential number of long-range, multi-body couplings is not a scalable protocol: if they are known for a certain number $d$ of qubits, due to the 
peculiar behaviour of the primes one cannot expect to use this knowledge to fix the couplings for the $(d+1)$ qubit system. 

Let's now look at the implementation of our algorithm adopting the \textit{analog} route in terms of Schr\"{o}dinger Hamiltonians. This is 
similar to what has been done recently \cite{holographic} for a quantum potential of the primes. Hence, we will deal with Hamiltonians of a particle of mass $m$ 
of the form 
$\hat{H}\,=p^2/2 m +V(x)$ 
with their eigenvalues determined by the Schr\"odinger equation $\hat{H}\psi_n=e_n \psi_n$.

A potential ${\cal V}_N(x)$ entering a Schr\"{o}dinger Hamiltonian with a finite and assigned
number of eigenvalues $\{e_0,e_1,...,e_M\}$ can be constructed using methods of Supersymmetric Quantum Mechanics (SQM)  \cite{SUSYKhare,QMprime1,QMprime2}. The procedure consists of solving iteratively the non-linear differential equations for the sequence of superpotentials $W_k(x)$
\begin{equation}
    W_k'(x)-W^2_k(x)+{\mathcal V}_{k-1}(x)=\tilde{E}_k \quad\quad , \quad \quad k=1,...,M \,\,\,, 
    \label{eq:Riccati}
\end{equation}
where $\tilde{E}_k=e_{M-k}-e_M$ are the negative spacings computed from the highest eigenvalue $e_M$, where the potential ${\cal V}_k(x)$ is obtained iteratively from
\begin{equation}
    {\cal V}_k(x)=2\tilde{E}_k+2W^2_k(x) - {\cal V}_{k-1}(x) \,\,\,.
    \label{eq:VfromW}
\end{equation}
The initial condition is set to be $W_k(0)=0$ as well as ${\cal V}_0(x)=0$ so that $e_0=0$. Substituting iteratively the ${\cal V}_k(x)$ of (\ref{eq:VfromW}) in (\ref{eq:Riccati}), one gets a system of nonlinear differential equations for the $W_k$. The potentials ${\cal V}_k(x)$ for $k=1,...,N-1$ are just intermediate steps of the computation since the final potential ${\cal V}_M(x)$, to be determined iteratively in $M$ steps.  
which accommodates all energy eigenvalues can be constructed using only $W_k(x)$. 
This procedure can be easily implemented to compute the potential $V(x)={\cal V}_M$ such that $\hat{H}=p^2/2m+V(x)$ has as eigenvalues the set of assigned values $\{e_0,e_1,...,e_M\}$. For instance, such a set can be of the first $M$ primes \cite{holographic,QMprime2} or the logarithm of the first $M$ primes \cite{mack10}. 

To conclude the implementation of our algorithm, let's then define the potentials $V_1(x)$ and $V_2(y)$ entering the two Schr\"{o}dinger Hamiltonians $\hat{H}_1$ and $\hat{H}_2$ of the previous section\footnote{It is worth to recall that the both potentials $V_1(x)$ and $V_2(y)$ can be also constructed in a semi-classical approximation\cite{GMscattering} for {\em all} primes and integers.} 
\be
\hat{H}_1(x)  \,=\, \frac{p_x^2}{2 m} + V_1(x) 
\,\,\,\,\,\,\,
,
\,\,\,\,\,\,\,
\hat{H}_2(y) \,=\, \frac{p_y^2}{2 m} + V_2(y) \,\,\,,
\label{H1H2}\ee
where $p_x$ and $p_y$ are the $x$ and $y$ components of its momentum, and such $\hat{H}_1$ ($\hat{H}_2$) has as eigenvalues 
the logarithm of the first $2^d$ primes (respectively, the logarithm of the first $2^d$ integers). The plot of $V_1(x)$ and $V_2(y)$ in a specific case is reported in SM 4.
Incidentally, if the Goldbach conjecture were true, 
taking the same Hamiltonian $\hat{H}_1$ in both the $x$ and $y$ direction, i.e. ${\cal H} = \hat{H}_1(x)+\hat{H}_1(y)$, the spectrum of ${\cal H}$ would be given by the logarithms of {\em all} even numbers.


\vspace{3mm}
\noindent
\textit{Discussion on computational complexity. }  
Once the potentials described above are experimentally realised, one can use the two 
Hamiltonians $\hat{H}_1$ and $\hat{H}_2$ to build up the Hamiltonian $\hat{H} = \hat{H}_1 + \hat{H}_2$ and proceed to factorize the integers $N$ below the cut-off $M$ 
through a sequence of $\hat{H}_1$ measurements. 
So, for example, if $N$ is a product of $3$ primes, only 
three quantum measurements are needed.
One can show that $k$ is clearly the lower bound of operations to factorize an integer, independently from quantum mechanics. Indeed, suppose that the integer $N$ has the form \eqref{primedecomposition1} and that one is given the information that $p_1,\cdots,p_k$ are its factors. To verify it, one divide by one of them, say $p_1$, and continue by it till it is possible ($\alpha_1+1$ divisions), and so on, for a total of $\sum_{i=1}^{k}\alpha_i+k$ divisions. If one is also provided the information about the multiplicities, then only $k$ divisions are needed. In the latter case, the last division is trivial.

The procedure proposed here saturates such lower bound. Indeed, from the first quantum measurement one gets one of the factors, and then divide 
$N$ 
(e.g. using a classical computer) by the latter till the obtained number is not longer divisible. Iterating this way one needs $k$ quantum measurements (the last, $k$-th, being equal to the identity) and $\sum_{i=1}^{k} \alpha_i$ divisions on the classical computer. One sees that the factorization based on quantum measurements 
saturates the lower bound, whether the multiplicities are given or they are not.


There is however a question to face: if the Shor algorithm has a complexity scaling with $O(\log N)$, i.e. of the number of digits, how 
how reconciled or put in relation 
with the fact that 
our scheme has a complexity of  
order $O(k)$, i.e. of the number of factors of $N$? As shown in the following, the answer is interesting and 
related to how the algorithm can be run. 
The difference $O(\log N)$ vs. $O(k)$ is emerging if one has a device in which the quantum measurement of an operator with assigned spectrum is physically possible. But, if such device is not at hand, and since one needs an operator having as eigenvalues the logarithms of the primes and another with the logarithms of the integers, one physically needs a device in which such operators with assigned spectra are implemented. The following discussion will show that -- if this device is not given and has to be implemented from scratch -- one is subjected
again to a complexity which is essentially $O(\log N)$, with some extra features discussed below. 

One has to realise that  
the determination of a final potential such as $V_M(x)$ with $M$ energy eigenvalues $e_k$ ($k=1,2,\ldots M$) has an algorithm complexity equal to $M$, because this is the number of differential equations to solve. Hence, in the presence of an exponential number of energy levels, there will be an exponential number of differential equations to solve to accommodate these levels. If this would be true, even though at a software level our algorithm has a finite degree of complexity, at its hardware level (i.e. to physically build the device) its complexity would grow exponentially.  
However, there is a very interesting way to cure the rapid exponential growth of complexity at the hardware level. The remedy comes from a 
result by Lloyd et al. \cite{lloyd} who showed that a system of $M$, possibly non-linear, differential equations of the form 
\begin{equation}
    \frac{d\vec{W}}{dx}+\mathbf{f}(W)\cdot \vec{W}=\vec{b}
    \label{FormLloyd}
\end{equation}
where $\mathbf{f}$ is a $M\times M$ matrix and $\vec{b}$ is a constant vector, can be solved on a quantum computer in $\log M$ number of steps! Namely, there is an 
exponential speed up which turns our problem to adjust an exponential number $2^d$ of eigenvalues in each of the Hamiltonians $H_1$ and $H_2$ in terms of only $d$ computational steps. 

Remarkably enough, the system of differential equations from supersymmetric quantum mechanics can be recast exactly as in eq.\,(\ref{FormLloyd}) by taking $\vec{W}=(W_1,...,W_M)^T \in \mathbb{R}^M$, and the entries of the (triangular) matrix $\mathbf{f}$ and the vector $\vec{b}$ given by  
\begin{equation}
    f_{ab} \,=\, \left\{
    \begin{array}{llll}
    0 & , & \,\, & a > b \\
    -W_a & , & \,\, & a =b \\
    (-1)^a 2 W_b & , & \,\,& a < b 
    \end{array}
    \right.
    \,\,\,\,\,\,\,\,\, 
    , 
    \,\,\,\,\,\,\,\,\,
    b_k = \tilde{E}_k+\sum^{k-1}_{i=1}(-1)^{i}2\tilde{E}_{k-i} \,\,\,.
    \end{equation}
Notice that, in the digital implementation of our algorithm, where $\hat{H}_1$ is a spin Hamiltonian of $d$ quantum spins, one has to solve an exponential number of linear algebraic equations. This ensures that the eigenvalues are the first $2^d$ primes, or their logarithms. So, one could use the quantum computer matrix inversion algorithm to solve them in order of $d$ steps. The point is that one has to physically implement an exponential number of multi-body, arbitrarily long-range couplings among the spins -- which must be contrasted with the analog implementation of the potentials $V_1$ and $V_2$, where the only limitation is provided by the resolution we can reach to realize them.

We conclude that the exponential growth of complexity at the hardware level for our procedure can be cured in digital form by 
using a (true) quantum computer, when available. This argument shows that 
the factorization protocol in terms of quantum measurements presented here and the Shor algorithm have similar complexity of $O(\log N)$. 
However, comparing in more detail the two algorithms, their pros and cons are the following: the Shor algorithm could run on a multi-purpose 
quantum computer and will take each time $O(\log N)$ steps to factorize a number $N$. 
On the contrary, our algorithm could run on a dedicated (not multi-purpose) device, 
which can be set up by solving, once for all, on a quantum computer a system of differential equations in $O(\log N)$ steps. 
Once this dedicated device has been set up, it perform factorization an integer $N$ in $k$ steps, $k$ being the number of prime factors, which is the least possible number of steps. 
It would be interesting to see if such devices can be used to solve other cryptographic and number theory problems in the future.


\vspace{3mm}
\noindent
{\bf Acknowledgments --} The authors thank D. Bernard, M. Berry, J. L. Cardy, D. Cassettari, A. Schwimmer and A. Smerzi for useful discussions and correspondence. 
GM acknowledges the grants PNRR MUR Project PE0000023-NQSTI 
and PRO3 Quantum Pathfinder. AT acknowledges MIT for kind hospitality during the writing of the final part of the manuscript. He also acknowledges the MIT-FVG Seed Fund Collaboration Grant "Non-Equilibrium Thermodynamics of Dissipative Quantum Systems". 
\vspace{3mm}

\newpage 

\appendix{\bf{Supplementary Material 1 - Preparation of the initial state}} 

To prepare the state $|\log N \rangle$ one can also employ the quantum measurements' projective nature. 
Namely, one can start with a state $|\Psi \rangle$ belonging to an ensemble with the expectation value of the energy equal to $\log N$: $\langle \Psi | \hat{H} | \Psi \rangle=\log N$ and variance $\Delta$. If $\Delta$ is quite narrow, the expansion of this state in eigenstates of $H$ will involve only a few terms nearby the eigenvalues $\log N$, i.e.
\begin{equation}
\mid \Psi \rangle\,=\, \sum_{\tilde N \in \{N - \Delta,N + \Delta\}} \gamma_n \, \mid \log \tilde N \rangle\,\,\,. 
\label{slowneutron}
\end{equation}
If we measure $\hat{H}$ on this state,  the measurement causes the system collapse on one of the eigenstates of $\hat{H}$ entering (\ref{slowneutron}). If the measurement output is $\log N$, we can start the procedure described above. If not, we re-prepare the state and re-measure $\hat{H}$ since, sooner or later the expected value will appear. Clearly, the number of iterations 
depends on the variance of $\hat{H}$ on $\Psi$. Another strategy could be 
to measure $\hat{H}_1$ on $\Psi$ and verify whether the state collapses on a factor of the integer $N$.

\vspace{5mm}

\appendix{\bf{Supplementary Material 2 - Examples of $\hat{H}_1$}} 

In both digital and analog versions of the algorithm presented in the main text, the requirements are: {\textit a)} the implementation of an operator $\hat{O}$ sum of two operators $\hat{O_1}$ and $\hat{O}_2$ having as eigenvalues, respectively, the logarithms of the prime numbers and of the integer numbers; and {\textit b)} the possibility to perform quantum measurements of $\hat{O}_1$.

Let's discuss the digital setting, referring to the main text for the analog case. We assume then that we have two registers, denoted by $1$ and $2$, of $d$ qubits each, labeled by the index 
$j=1, \cdots, d$. The observable $\hat{O}_1$ acts only on register $1$, and similarly 
$\hat{O}_2$ acts only on register $2$, where the first operator has, as eigenvalues, the (natural) logarithms of the first $2^d$ primes starting from prime $2$, while  
the second operator has, as eigenvalues, the (natural) logarithms of the first $2^d$ natural numbers. 


To fix the notation, let discuss the operators $\hat{\cal O}_1$ and $\hat{\cal O}_2$ having as eigenvalues respectively the first $2^d$ prime number and the first $2^d$ integers. Formally one can then implement the desired operators $\hat{O}_1$ and $\hat{O}_2$, since the former (latter) has the same eigenvalues  of $\hat{\cal O}_1$ and $\hat{\cal O}_2$.

To build $\hat{\cal O}_2$ is  
considerably easier -- actually, exponentially easier than to construct $\hat{\cal O}_1$.   
First of all, let's consider in the second register the operator \cite{Nielsen2010} 
 $$\hat{Z}_j= \mathbb{1} \otimes \mathbb{1} \otimes \cdots \otimes 
 \stackrel{\stackrel{j}{\downarrow}}
 { Z} \otimes \cdots \mathbb{1}$$
 where 
$$\mathbb{1}=\begin{pmatrix}
    1 & 0\\
    0 & 1
    \end{pmatrix}
 \,\,\,\,\,\,\,\,\,
 ,
 \,\,\,\,\,\,\,\,\,
 Z=
    \begin{pmatrix}
    1 & 0\\
    0 & -1
    \end{pmatrix}
 $$
and the symbol $\otimes$ denotes the tensor product between matrices 
\footnote{We remind for convenience that if one has two matrices $A_{ij}$ and $B_{i'j'}$, then 
the matrix $A \otimes B$ has as matrix elements $(A \otimes B)_{ii',jj'}=A_{ij} \cdot B_{i'j'}$, so that, e.g., $Z \otimes \mathbb{1}$ is the $4\times4$ matrix 
 $\begin{pmatrix}
    \mathbb{1} & \mathbb{0}\\
    \mathbb{0} & -\mathbb{1}
    \end{pmatrix}$ written as a block matrix,
    where $\mathbb{0}$ is the $2 \times 2$ matrix $\mathbb{0}=   \begin{pmatrix}
    0 & 0\\
    0 & 0
    \end{pmatrix}
$.}. 
It is then clear that the $\hat{Z}_j$ are $2^d \times 2^d$ matrices and correspond to the observable $z$-component of the $j$-th spin \cite{Sakurai2020}. Let's now define the $2 \times 2$ matrix ${\cal N} =(\mathbb{1}+Z)/2 =   \begin{pmatrix}
    1 & 0\\
    0 & 0
    \end{pmatrix}
    $,
having evidently $0$ and $1$ as eigenvalues, and the corresponding matrices 
$$\hat{{\cal N}}_j=\mathbb{1} \otimes \mathbb{1} \otimes \cdots \otimes 
\stackrel{\stackrel{j}{\downarrow}}{\cal{N}}
 \otimes \cdots \mathbb{1}\,\,\,.$$ 
 One can immediately show that 
the operator
\begin{equation}
    \hat{\cal O}_2=2 \, \hat{\mathbb{1}}_d+\sum_{j=1}^{d} 2^{j-1} \hat{{\cal N}}_j 
\label{OO2}
\end{equation}
has as eigenvalues the first $2^d$ integers starting from $2$, with $\hat{\mathbb{1}}_d=\mathbb{1} \otimes \mathbb{1} \otimes \cdots \otimes \mathbb{1}$ being the $2^d\times 2^d$ identity matrix. The rationale is that the sum $\sum_{j} 2^{j-1} \hat{{\cal N}}_j$ in \eqref{OO2} gives the binary representation of the numbers from $0$ to $2^d-1$. For example, for $d=3$, the operator 
$2^0 \hat{N}_1+2^1 \hat{N}_2+2^2 \hat{N}_3$ has the eigenvalues $0,1,2,3,\cdots,7$, as one can immediately verify. 
So it remains to sum to \eqref{OO2} only the constant $2$.

Let now come to $\hat{{\cal O}}_1$. A strategy to give $\hat{{\cal O}}_1$ in terms of the operators $\hat{{\cal N}}_i$'s 
is to write all possible multi-spin operators in which the $\hat{{\cal N}}_i$ enter $0$ times, $1$ times, $2$ times, $3$ times and so on. Borrowing the notation from treatments of one-dimensional spin chains with multi-spin interactions, see e.g. \cite{Gori2011}, one can write
$$
   \hat{\cal O}_1=j_\emptyset \, \hat{\mathbb{1}}_d + \sum_{i_1} j_{i_1}\hat{N}_{i_1} 
    +\sum_{i_1,i_2} j_{i_1,i_2} \hat{N}_{i_1} \hat{N}_{i_2} +
$$
\begin{equation}
    +\sum_{i_1,i_2,i_3} j_{i_1,i_2,i_3} \hat{N}_{i_1} \hat{N}_{i_2}\hat{N}_{i_3}
    +\cdots+j_{i_1,i_2,i_3,\cdots,i_m} \hat{N}_{i_1}\hat{N}_{i_2} \cdots \hat{N}_{i_m} \,,
\label{OO1bis}
\end{equation}
with all sums run from $1$ to $d$. Now the statement is that one can choose the coefficients $\left\{j_\mu\right\}$ 
such that the the eigenvalues of $\hat{\cal O}_1$ are exactly the first $2^d$ primes, since the total number of the coefficients $\left\{j_\mu\right\}$ is exactly  $2^d$: to do so, 
one has to solve a linear system of $2^d$ equations in the unknowns $\left\{j_\mu\right\}$ using the first $2^d$ primes are the input.

For instance, in the case  $d=3$, with the values of the coefficients given by 
$$j_\emptyset =2 \,\,, \,\, j_{1}=1 \,\,,\,\, j_{2}=3\,\,,\,\, j_{3}=5\,\,, \\
j_{12}=5\,\,,\,\, j_{23}=3\,\,,\,\, j_{13}=9\,\,,\,\, j_{123}=-9$$ 
one gets the first $2^3=8$ primes $2,3,5,7,11,13,17,19$.

Up to $d=5$, the explicit expressions of $\hat{\cal O}_1^{(d)}$ having the first $2^d$ primes starting from $p=2$ and 
defined on $d$ qubits are given by 

\begin{eqnarray*}
\hat{\cal O}_1^{(2)} &=& 2 \hat{\mathbb{1}} + 1 \hat{N}_1 + 3 \hat{N}_2 + 1 
\hat{N}_1 \hat{N}_2 \\
\hat{\cal O}_1^{(3)} &= &2 \mathbb{1}_3 + 1 \hat{N}_1 + 3 \hat{N}_2 + 5 \hat{N}_3 + 5 \hat{N}_1 \hat{N}_2 + 3 \hat{N}_2 \hat{N}_3 + 9 \hat{N}_1 \hat{N}_3 - 9  \hat{N}_1 \hat{N}_2 \hat{N}_3 \\
\hat{\cal O}_1^{(4)} &=& 2 \hat{\mathbb{1}}_4+ 1 \hat{N}_1 + 3 \hat{N}_2 + 5 \hat{N}_3 + 9 \hat{N}_4 + 7 \hat{N}_1 \hat{N}_2 + 9 \hat{N}_1 \hat{N}_3+ 
   7 \hat{N}_1 \hat{N}_4 + 13 \hat{N}_2 \hat{N}_3 + \\ 
   & + & 15 \hat{N}_2 \hat{N}_4 + 15 \hat{N}_3 \hat{N}_4 - 3 
   \hat{N}_1\hat{N}_2\hat{N}_3 - 
   3 \hat{N}_1 \hat{N}_2 \hat{N}_4 - 5 
   \hat{N}_1 \hat{N}_3 \hat{N}_4 - 15 
   \hat{N}_2 \hat{N}_3 \hat{N}_4- 
   7 \hat{N}_1 \hat{N}_2 \hat{N}_3 \hat{N}_4\\
  \hat{\cal O}_1^{(5)} & = & 2 \mathbb{1}_5 + 1 \hat{N}_1 + 3 \hat{N}_2 + 5 \hat{N}_3 + 9 \hat{N}_4 + 11 \hat{N}_5 + 11 \hat{N}_1 \hat{N}_2 + 
  11 \hat{N}_1 \hat{N}_3 + 11 \hat{N}_1 \hat{N}_4 + 15 \hat{N}_1 \hat{N}_5 +\\
  &+ & 21 \hat{N}_2 \hat{N}_3 + 23 \hat{N}_2 \hat{N}_4 + 
  25 \hat{N}_2 \hat{N}_5 + 27 \hat{N}_3 \hat{N}_4 + 29 \hat{N}_3 \hat{N}_5 + 31 \hat{N}_4 \hat{N}_5 +
  5 \hat{N}_1 \hat{N}_2 \hat{N}_3 + \hat{N}_1 \hat{N}_2 \hat{N}_4 
  -   \hat{N}_1 \hat{N}_2 \hat{N}_5 + \\
  &+& 5 \hat{N}_1 \hat{N}_3 \hat{N}_4 - 
   \hat{N}_1 \hat{N}_3 \hat{N}_5 
  -1 \hat{N}_1 \hat{N}_4 \hat{N}_5 - 7 \hat{N}_2 \hat{N}_3 \hat{N}_4 -
  7 \hat{N}_2 \hat{N}_3 \hat{N}_5 - 7 \hat{N}_2 \hat{N}_4 \hat{N}_5 
   -  13 \hat{N}_3 \hat{N}_4 \hat{N}_5 + \\
   &-& 
  25 \hat{N}_1 \hat{N}_2 \hat{N}_3 \hat{N}_4 - 23 \hat{N}_1 \hat{N}_2 \hat{N}_3 \hat{N}_5 - 25 \hat{N}_1 \hat{N}_2 \hat{N}_4 \hat{N}_5 - 
  29 \hat{N}_1 \hat{N}_3 \hat{N}_4 \hat{N}_5 - 25 \hat{N}_2 \hat{N}_3 \hat{N}_4 \hat{N}_5 + 49 \hat{N}_1 \hat{N}_2 \hat{N}_3 \hat{N}_4 \hat{N}_5
  \end{eqnarray*}
   
E.g., one can directly verify that the eigenvalues of $\hat{\cal O}_1^{(5)}$ are the first $2^5=32$ primes 
  $$ p =2,3,5,7,11,13,17,19,23,29,31,37,41,43,47,53,59,61,
  71,73,79,83,89,97,101,103,107,109,113,127,131$$

In the course of our analysis we observe the emergence of some regularity: first of all that the coefficients are relative integers  
moreover taht $j_\emptyset=2$, $j_i=p_i-2$, where $p_i$ is the $i$th prime and, finally, that increasing the number of qubits implies that the $j_\mu$ pass form positive to negative values. But, at the same time, we notice that there is no simple pattern of the remaining coefficients $\left\{j_\mu\right\}$: if one has the coefficients  $\left\{j_\mu\right\}$ for a certain value of $d$, it seems that there is no a simple procedure to get the $\left\{j_\mu\right\}$ relative to $d+1$. After all this is not surprising, since we know that even very effective ways to plot the primes can display striking regularities (see, for instance, the pattern emerging from Ulam spirals\cite{Ulam}), but they cannot predict the $d$-prime simply by knowing the first $d-1$ primes. In other words, to determine for each $d$ the coefficients $\left( j_\mu \right\}$ one has to solve a system with $2^d$ unknowns assuming that we {\it know} the first $2^d$ primes.


A way to visualize the coefficients $\left\{ j_\mu \right\}$ is to put in a table like the following, where the coefficients are written in the form $j_{i_1,\cdots,i_k}$ with $i_1 < i_2 <\cdots < i_k$, ordering them in a way that among two $k$ tuples $(i_1,\cdots,i_k)$ 
and $(j_1,\cdots,j_k)$, the one with the smallest among $i_1$ and $j_1$ is listed before, and if $i_1=j_1$ then the one with the smallest among $i_2$ and $j_2$ is listed before, and so one. Therefore, for $k=3$ and $m=5$, we report in the table the values of $j_{123}$, $j_{124}$, $j_{125}$, $j_{134}$, $j_{135}$, 
  $j_{145}$, $j_{234}$, $j_{235}$, $j_{245}$, $j_{345}$.  
\vspace{3mm}  
\begin{widetext}
\begin{center}
\begin{tabular}{ |c|c|c|c|c|c| } 
 \hline
 m $\to$ & 2 & 3 & 4& 5\\ 
 \hline \hline
$j_\emptyset$ & 2 & 2 & 2 & 2 \\ 
  \hline
$ j_{i} $&  1, 3 & 1,3,5  & 1,3,5,9 & 1,3,5,9,11\\
  \hline
$ j_{ij} $& 1 & 5, 3, 9   &7, 9, 7, 13, 15, 15  &  11, 11, 11, 15, 21, 23, 25, 27, 29, 31\\
  \hline
$  j_{ijk} $ & 0 & -9  & -3,-3,-5,-15 & 5,1,-1,5,-1,-1,-7,-7,-7,-13 \\
  \hline
$ j_{ijkl} $ & 0 & 0 & -7 & -25,-23,-25,- 29,-25\\
  \hline
$ j_{ijklm} $ & 0 & 0 & 0 & 49\\ 
 \hline
\end{tabular}
\end{center}
\end{widetext}
\vspace{3mm}
Similarly to what we described above, one could determine the couplings $j$'s in such a way that $\cal{O}_1$ has as eigenvalues the logartithm of the first $2^d$ primes: in the few attempts we have done, their values appear erratic as well. 

\vspace{5mm}

\appendix{\bf{Supplementary Material 3 - Gedankenexperiment}}\label{gedanken}

In this section we discuss a Gedankenexperiment for implementing the measurement procedure discussed in Section 2. First of all, let's imagine that the Hamiltonians $H_1$ and $H_2$ are realised, as in \cite{holographic}, in terms of some laser optical device. We put together this optical device inside a metallic box $S$ and transfer an electric charge $Q$ on it so that this metallic box acts as a Faraday cage which isolates the quantum system inside the box from external perturbations.  From an outside, such a metallic box $S$ can  simply be characterised in terms of its electric charge $Q$ and its mass $M$. The mass $M$ is made by the mass $m_b$ of the metallic box and the quantum energy levels occupied by the particle, assuming the equivalence between masses and energies (in the following we take the units such that the speed of light is equal to 1, $c =1$). Being the Hamiltonian $H$ made of two commuting Hamiltonians $H_1(x)$ and $H_2(y)$ along the axes $x$ and $y$, the system the have two different inertias regarding the two axes: if $e^{(1)}_i$ and $e^{(2)}_j$ denote the $i$-th and the $j$-th energies along the $x$ and the $y$ axis, there will be an horizontal mass $M_x = m_b + e^{(1)}_i$ and a vertical mass $M_y = m_b + m^{(2)}_k$. 

\begin{figure}[t]
 \includegraphics[width=0.4\textwidth]{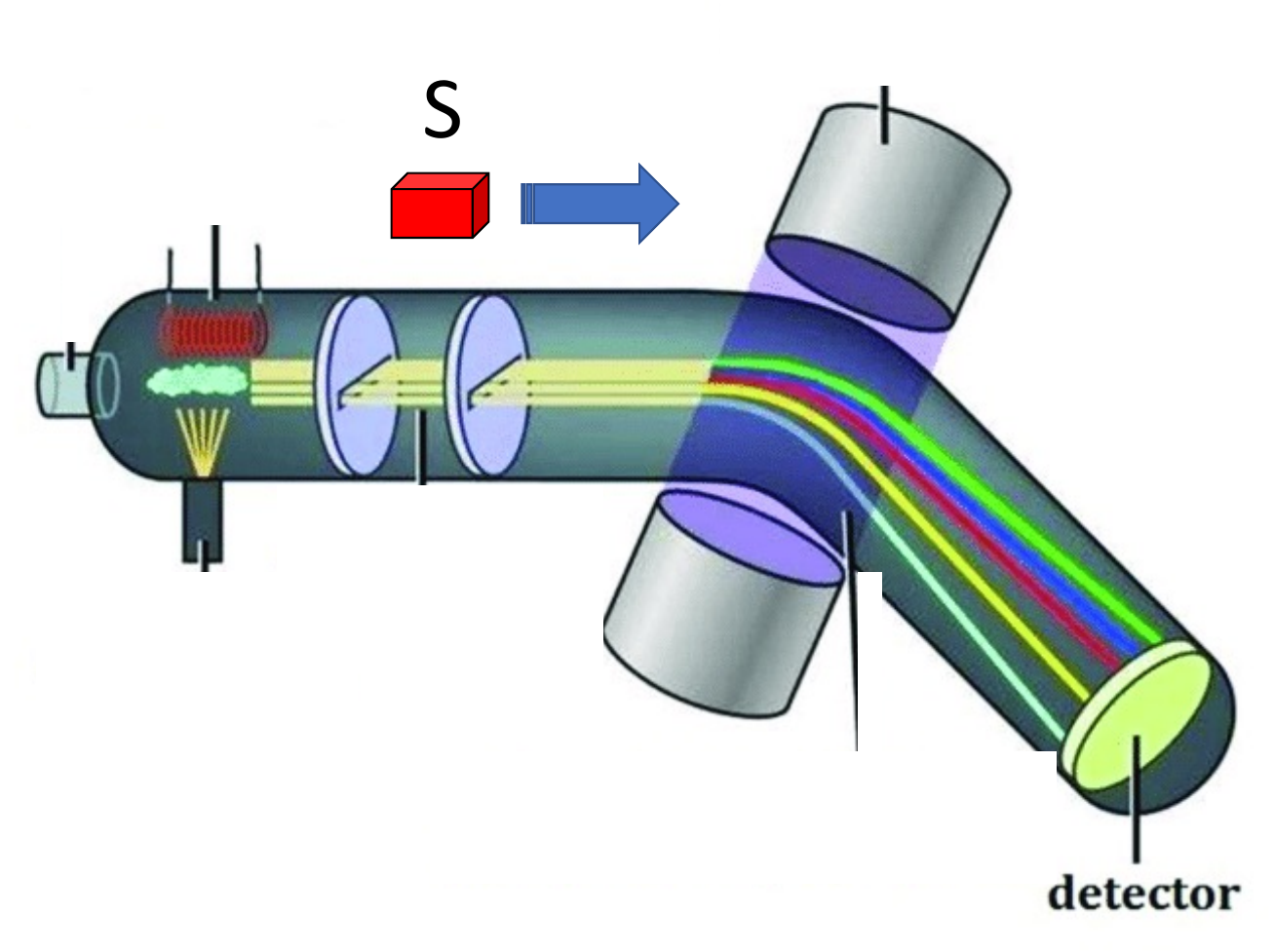}
    \caption{Mass spectrometer which permits to measure the horizontal mass $M_x$ of the system $S$ and therefore to project on one of the horizontal energy eigenstate 
    $| e^{(1)}_a\, \rangle$.}
        \label{spectrometer}
\end{figure}

We can send our system $S$ to a mass spectrometer, such as the one shown in Figure \ref{spectrometer}. Such a classical apparatus will set up a Stern-Gerlach experiment and 
 we can measure the energy projectively. Indeed, skipping the details of the motion, the system $S$ trajectory is determined by the ratio $Q/M_x$. Hence, if the system $S$ is in a linear superposition of $k$ energy eigenstates of $H_1(x)$ (as it is indeed the case for our vector (\ref{genericstate}))
\be
| \psi \,\rangle \,=\, \sum_{i=1}^k a_i | e^{(1)}_i\,\rangle  \,\,\,,
\label{psis}
\ee
then, when the system $S$ passes through the mass spectrometer, it will end up in {\em one} of the $k$ different trajectories $\{T_1, T_2, \ldots T_k\}$, as shown in Figure 
\ref{spectrometer}. Imagine that the system ends following the $a$-th trajectory, then after the passage through the mass spectrometer, the system is projected on the energy eigenstate $| e^{(1)}_a \, \rangle$.  

\vspace{5mm}
\appendix{\bf{Supplementary Material 4 - Potentials $V_1$ and $V_2$}}\label{gedankenn}

\begin{figure}[t]
 \includegraphics[width=0.71\textwidth]{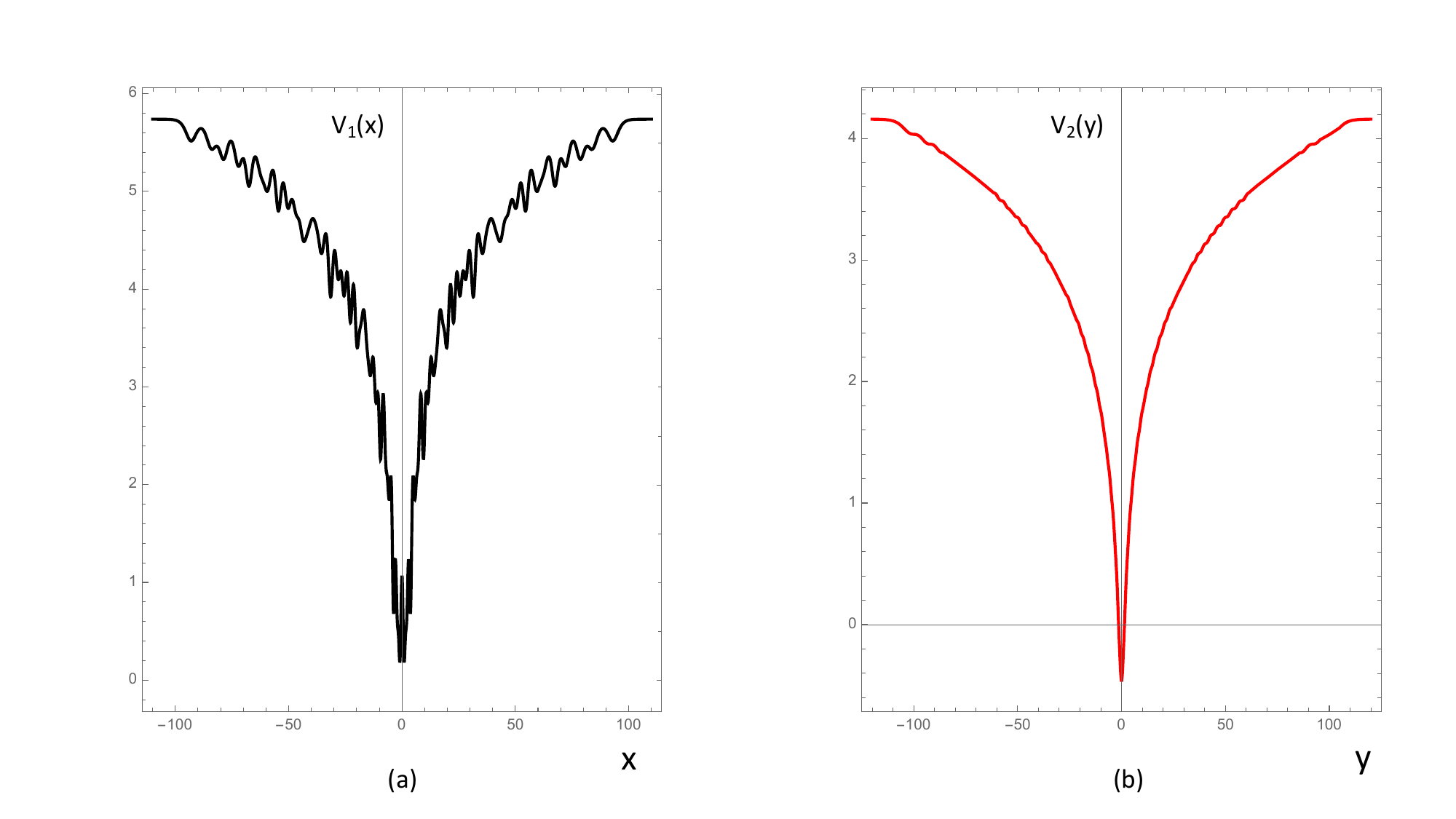}
    \caption{Plots of the potentials $V_1(x)$ (a) and $V_2(y)$ (b) relative to $N=64 = 2^5$  energy levels given respectively by the logarithm of the primes, $\log p_k$,  and 
    the logarithm of the integers, $\log m$. }
    \label{potentials}
\end{figure}

In Figure \ref{potentials} we show the plots of $V_1(x)$ and $V_2(y)$ with 
a spectrum given respectively by $\log p_k$ and $\log m$ for the first $M = 64 =2^6$ energy levels of each potential. 

\end{document}